\documentclass[doublecol]{epl2}

\usepackage{graphicx}
\usepackage{dcolumn}
\usepackage{bm}
\usepackage{amstext}
\usepackage{color}

\begin{document}

\title{Low-Reynolds number swimming in gels}
\author{Henry C. Fu \and Vivek B. Shenoy \and Thomas R. Powers}
\institute{Division of Engineering, Brown University, Providence, RI 02912}

\abstract{ Many microorganisms swim through gels, materials with
  nonzero zero-frequency elastic shear modulus, such as mucus.
  Biological gels are typically heterogeneous, containing both a
  structural scaffold (network) and a fluid solvent.  We analyze the
  swimming of an infinite sheet undergoing transverse traveling wave
  deformations in the ``two-fluid'' model of a gel, which treats the
  network and solvent as two coupled elastic and viscous continuum
  phases.  We show that geometric nonlinearities must be incorporated
  to obtain physically meaningful results.  We identify a transition
  between regimes where the network deforms to follow solvent flows
  and where the network is stationary.  Swimming speeds can be
  enhanced relative to Newtonian fluids when the network is
  stationary.  Compressibility effects can also enhance swimming
  velocities.  Finally, microscopic details of sheet-network
  interactions influence the boundary conditions between the sheet and
  network.  The nature of these boundary conditions significantly
  impacts swimming speeds.  }

\maketitle

While biological locomotion at low-Reynolds number is a
well-established and active field (see~\cite{LaugaPowers2009} for a
review), only recently has swimming in complex, non-Newtonian media
started to be systematically explored.  Indeed, many microorganisms
routinely swim through complex or non-Newtonian media.  Bacteria in
the stomach such as {\em Helicobacter pylori} encounter gastric mucus~\cite{Celli_etal2009};
mammalian sperm swim through viscoelastic mucus in the female
reproductive tract~\cite{FauciDillon2006,SuarezPacey2006}; and spirochetes burrow into the tissues they
infect~\cite{Canole-Parola1978}.  Most of the prior theoretical work on swimming in complex
media has focused on viscoelastic {\em fluids}~\cite{Chaudhury1979,
  FulfordKatzPowell1998, Lauga2007, FuPowersWolgemuth2007,
  FuWolgemuthPowers2008Shapes, FuWolgemuthPowers2008Scallop,
  NormandLauga2008, Lauga2009, TeranFauciShelley2009}.  Because
prescribed swimming strokes lead to the same swimming speed in
linearly viscoelastic fluids as in Newtonian
fluids~\cite{FulfordKatzPowell1998, Lauga2009}, these studies have
focused on the effects of nonlinear
viscoelasticity~\cite{Chaudhury1979, Lauga2007, FuPowersWolgemuth2007,
  FuWolgemuthPowers2008Scallop, NormandLauga2008, Lauga2009,
  TeranFauciShelley2009} and the effect of altered medium response on
swimming strokes via fluid-structure
interactions~\cite{FuPowersWolgemuth2007,
  FuWolgemuthPowers2008Shapes}.  However, many biological viscoelastic
materials contain crosslinked polymer networks, and are therefore
solid rather than fluid.  In this paper we focus on swimming in such
crosslinked materials, which have a nonzero zero-frequency elastic
shear modulus.  We use the term ``gel'' to refer to these types of
materials.

Generally, viscoelastic materials have frequency-dependent response,
while the distinction we make between solid and fluid is a
zero-frequency property.  Some of the issues addressed by studies of
viscoelastic fluids carry over straightforwardly to gels.  For
example, the fluid-structure interactions which determine the beating
patterns of sperm flagella depend on the viscoelastic response at the
beating frequency, not at zero frequency.  However, a simple physical
argument demonstrates a fundamental difference between swimming in a
gel versus swimming in a fluid.  In fluid dynamics, one uses no-slip
boundary conditions, but for a swimmer in a solid,
no-slip boundary conditions do not allow swimming.
Under no-slip boundary conditions a swimmer would drag solid
material along with its motion, leading to large deformations and
restoring forces which oppose translation.  Thus, swimming in a gel
requires that the swimmer slips past the medium.  This simple
consideration highlights the importance of boundary conditions for
understanding swimming in gels, a major theme of this
paper.

Another motivation for studying swimming in gels is the search for
mechanisms that enhance swimming speeds in complex materials.  Due
to the observation that spirochetes can swim faster in fluids with an
elastic response component~\cite{BergTurner1979}, there has been an
interest in how viscoelastic media can increase or decrease swimming
velocities.  For filamentous geometries, it has been suggested
that polymeric fluids may have higher effective anisotropies in
resistive-force theories, leading to increased swimming
speeds~\cite{MagariyamaKudo2002, Leshansky2009}.  Furthermore, the
breaking of kinematic reversibility~\cite{happel_brenner1965} in
nonlinearly viscoelastic fluids makes certain reciprocal swimming
strokes more effective than in Newtonian
fluids~\cite{FuWolgemuthPowers2008Scallop, Lauga2009}.  Restricting
attention to sheet-like geometries undergoing traveling-wave
deformations (the subject of this paper), an infinite sheet in a
nonlinearly viscoelastic fluids swims more slowly than in a Newtonian
medium~\cite{Lauga2007}.  However, finite-length sheet-like swimmers
in viscoelastic fluids~\cite{TeranFauciShelley2009}, and infinite
swimmers in a Brinkmann fluid can show increased swimming
velocities~\cite{SiddiquiAnsari2003, Leshansky2009}.

The latter two studies~\cite{SiddiquiAnsari2003, Leshansky2009} are
closest in spirit to this letter.  Both investigate how Taylor's
swimming sheet~\cite{Taylor1951}, an infinite sheet with a small
amplitude traveling wave deformation, behaves in the Brinkmann model,
which treats the effect of a stationary porous phase on fluid flows in
an effective medium approach.  Instead, we solve for the flows induced
by the moving sheet in the ``two-fluid model'' description of a
gel~\cite{Levine_Lubensky2001a, deGennes1976a, Milner1993,
  FuShenoyPowers2008}, which represents the gel as two phases.
One phase describes an elastic polymer network with displacement field
$\mathbf{u}$ and Lam\'e coefficients $\mu$ and $\lambda$, and the
other phase describes a viscous fluid solvent with velocity field
$\mathbf{v}$ and viscosity $\eta$.  The two phases are coupled by a
drag force proportional to their relative local velocity, with drag
coefficient $\Gamma$, giving the governing equations
\begin{eqnarray}
\nabla \cdot {\bm \sigma}^\mathrm{n} &=&  \Gamma \left( \frac{\mathrm{d}}{\mathrm{d}t}{
\mathbf{u}} - \mathbf{v} \right) \label{gelequations1}\\
 \nabla \cdot {\bm \sigma}^\mathrm{s}  &=& -\Gamma \left( \frac{\mathrm{d}}{\mathrm{d}t}{
\mathbf{u}} - \mathbf{v} \right) \label{gelequations2}\\
\bm{\sigma}^{\mathrm{n}} &=& \mu \left[ \nabla \mathbf u
+ (\nabla \mathbf{u})^\mathrm{T} \right] + \mathbf{I} \; \lambda
\nabla \cdot \mathbf{u} + \bm{\bar \sigma}^{\mathrm{n}} \label{polymerstress}\\
\bm{\sigma}^{\mathrm{s}} &=& \eta \left[ \nabla \mathbf{v} + (\nabla \mathbf{v}
)^\mathrm{T} \right] - p \; \mathbf{I}\\
-\partial_t \varphi &=&  \nabla \cdot \left[
  \varphi \frac{\mathrm{d}}{\mathrm{d}t}\mathbf{u}\right] \label{conservation-u}\\
-\partial_t (1- \varphi) &=& \nabla \cdot \left[ (1 - \varphi)
  \mathbf{v} \right].\label{conservation-v}
\end{eqnarray}
Here the network and solvent stress tensors are
$\bm{\sigma}^\mathrm{n}$ and $\bm{\sigma}^\mathrm{s}$, respectively.
The moduli in the network stress tensor are effective, macroscopic
moduli -- for example, they incorporate osmotic effects, and they
depend implicitly on the volume fraction of the network.
Eq.~\ref{conservation-u} and \ref{conservation-v} express volume
conservation for each phase in terms of the network volume fraction
$\varphi$ assuming that the individual solvent and network phases are
incompressible. Combining the two yields
\begin{equation}
0 = \nabla \cdot \left[ (1 - \varphi) \mathbf{v} + \varphi
  \frac{\mathrm{d}}{\mathrm{d}t}\mathbf{u}\right].\label{conservation-full}
\end{equation}
In the rest of the paper, we
work in the regime where network volume fraction $\varphi$ can be
ignored, but yet the macroscopic network stresses cannot be ignored,
i.e. if we expand the solutions in $\varphi$, we obtain the
zeroth-order solutions in $\varphi$.  In this
$\varphi \rightarrow 0$ limit, Eq.~\ref{conservation-full} reduces to
$\nabla \cdot \mathbf{v} =0.$
The drag coefficient $\Gamma$ can be estimated as $\Gamma \approx
\eta/\xi^2$, where $\xi$ is the mesh size of the network~\cite{Levine_Lubensky2001a}.
Finally, note that the network stress tensor
incorporates nonlinear terms (involving products of
$\mathbf{u}$) explicitly in $\bar {\bm \sigma}^{\mathrm n}$.
 
Our use of the two-fluid model explicitly recognizes the role of
heterogeneity in complex media, while still retaining a continuum
description.  Ref.~\cite{Leshansky2009} emphasizes that
polymers in a viscoelastic fluid may be viewed as heterogeneous
inclusions in a viscous fluid.  Treating the inclusions as a
stationary background results in the Brinkmann fluid as an effective
model.  The two-fluid model allows us to describe
the heterogeneous phase more realistically as a dynamic component, and
understand when it is appropriate to think of heterogeneities as
stationary or dynamic.

Our investigation leads to three main conclusions.  First, network
dynamics determine when the presence of heterogeneities leads to
increased swimming velocities.  Roughly speaking, network deformations
are a result of drag forces arising from solvent flow, and are opposed
by the stiffness of the network.  For fast flows or compliant
networks, the networks move with the flows and do not attenuate the
flow, while for slow flows or stiff networks, the network acts like a
stationary phase, leading to enhancements of the swimming velocity.
We also find that compressibility of the network can lead to
additional enhancement of swimming velocities.

Second, a consistent treatment of swimming velocities requires proper
treatment of nonlinearities.  We show that nonlinearities in the
constitutive relationship ($\bar {\bm{\sigma}}^{\mathrm{n}}$ in
Eq.~\ref{polymerstress}) of the elastic phase of a gel do not affect
swimming speeds.  However, geometric nonlinearities arise from the
terms of Eqs.~\ref{gelequations1} and \ref{gelequations2} that involve the velocity of material
points of the network, $\frac{\mathrm{d}}{\mathrm{d}t} \mathbf{u}$.
The displacement field is a function of current coordinates -- i.e.,
the position of a material element originally at
$\mathbf{x}_0$ is $\mathbf{x}(\mathbf{x}_0,t) =
\mathbf{x}_0 + \mathbf{u}(\mathbf{x},t)$ -- so the velocity of
material elements is given by the material derivative, defined
implicitly via $\frac{\mathrm{d}}{\mathrm{d}t}
\mathbf{u} = \partial_t {\mathbf{u}} + \frac{\mathrm{d}}{\mathrm{d}t} \mathbf{u} \cdot \nabla
\mathbf{u}$.  
In the remainder of the paper we refer to the term
$\frac{\mathrm{d}}{\mathrm{d}t} \mathbf{u} \cdot \nabla \mathbf{u}$ as the
``convective'' nonlinearity.  Without it one cannot
obtain physically meaningful swimming velocities.

\begin{figure}
\includegraphics[width=8.5cm]{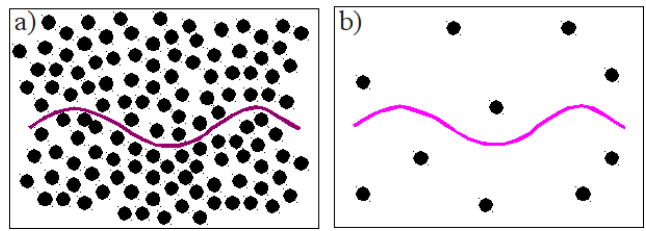}
\caption{ Effective sheet-network boundary conditions depend on the
  microscopic swimmer-network interactions.  a) Network is dense on
  the scale of swimmer deformations leading to direct network-sheet
  boundary conditions (Eqs.~\ref{direct1} and \ref{direct}).  b) Network is sparse on the scale of swimmer
  deformations leading to solvent-mediated network-sheet boundary
  conditions (Eq.~\ref{solvent}).  }\label{schematic}
\end{figure}

Finally, we highlight the importance of boundary conditions (BC)
between the sheet and elastic phase.  There is a long history of
experimental support for the validity of no-slip boundary conditions
for fluids, and we use no-slip BC between the sheet and solvent,
$\mathbf{v}(x,h(x,t),t) = \partial_t h(x,t) \hat{\bf{y}}$, where $h(x,t)$ is
height of the sheet at $(x,t)$.  However, there is no a priori
selection of boundary conditions between the sheet and network in our
macroscopic theory.  In this paper, we consider two types of
network-sheet BC that may plausibly result from microscopic
interactions at a swimmer surface.

The first class of boundary conditions applies when the network is
dense on the scale of the swimmer (Fig.~\ref{schematic}a).  Then we imagine
that the sheet presses {\em directly} against the polymer network and
the network must conform to the sheet.
Using reference positions for the network produced by a flat sheet,
\begin{equation}
u_y(x,h(x,t),t)  = h(x,t).\label{direct1}
\end{equation}
The remaining boundary condition is applied to the tangential network
stress via a kinetic friction law, with tangential stress proportional
to the tangential velocity of the polymer relative to the sheet:
\begin{eqnarray}
\hat{\bf t} \cdot \bm{\sigma}^{\mathrm n}(x,h(x,t),t) \cdot \hat{\bf
  n} =&& \nonumber\\ && \!\!\!\!\!\!\!\!\!\!\!\!\!\!\!\!\!\!\!\!\!\!\!\!\!\!\!\!\!\!\!  \Xi \hat{\bf
  t} \cdot \left( \frac{\mathrm{d}}{\mathrm{d}t} {\mathbf{u}}(x,h(x,t),t) - \partial_t h(x,t)
  \hat{\bf{y}} \right),\label{direct}
\end{eqnarray}
where $\hat{\bf t}$ and $\hat{\bf n}$ are the tangent and normal to
the sheet.  Note that $\Xi \rightarrow 0$ corresponds to
free tangential slip, while $\Xi \rightarrow \infty$
corresponds to no-slip.

In the second class of boundary condition, the network is dilute on
the scale of the swimmer (Fig.~\ref{schematic}b), so the sheet
exerts no forces directly on the network.  Instead all forces on the
network are mediated by the solvent.  Then the two components of the
network stress acting on the sheet must vanish:
\begin{equation}
\bm{\sigma}^{\mathrm{n}}(x,h(x,t),t) \cdot \hat{\bf n}  =0 \label{solvent}.
\end{equation}

Both the ``direct'' BC (Eqs.~\ref{direct1},\ref{direct}) and
``solvent-mediated'' BC (Eq.~\ref{solvent}) allow the swimmer to slide past the network, in accordance
with our physical discussion above.  

\textbf{Solutions}.  In the frame where the sheet is stationary, the
transverse displacements are described by $h(x,t) = \mathrm{Re}[ b
  \exp(i (k x - \omega t)]$.  Throughout this paper we
will be concerned with the zero-Reynold's number limit appropriate
for swimming microorganisms, and hence always ignore inertial
effects.

The sheet swims with velocity $V_s$, but sufficiently far from the
sheet in the lab frame the elastic phase is stationary.  Thus we write
the displacement field in the frame of the sheet as
$\mathbf{u}(\mathbf{x},t) = V_s \, t\, \hat {\bf x} + \delta
\mathbf{u}(\mathbf{x},t)$, with $\delta \mathbf{u}$ small.  Note that
$V_s$ turns out to be second order in the displacements.  The solution
strategy follows Taylor~\cite{Taylor1951}.  For small traveling wave
amplitudes relative to wavelengths, we determine the velocity and
displacement fields order by order in $b$.  We write (for example)
$\delta\mathbf{u} = \delta\mathbf{u}^{(1)} + \delta\mathbf{u}^{(2)} +
...$, with $\delta \mathbf{u}^{(1)}$ first order in $b$, etc.  The
fields satisfy the governing equations and boundary conditions, also
expanded order by order in $b$.



The first-order governing equations are
\begin{eqnarray}
\mu \nabla^2 \delta\mathbf{u}^{(1)} + (\lambda + \mu) \nabla(\nabla \cdot
\delta\mathbf{u}^{(1)})  &=& \Gamma \left( 
\partial_t \delta\mathbf{u}^{(1)} - \mathbf{v}^{(1)} \right) \nonumber
\\
\eta \nabla^2 \mathbf{v}^{(1)} - \nabla p &=&
-\Gamma \left( \partial_t \delta \mathbf{u}^{(1)} - \mathbf{v}^{(1)} \right).\nonumber
\end{eqnarray}
The first-order solutions take the form $\mathbf{v}^{(1)} = \mathrm{Re}
\lbrace \tilde\mathbf{v}(y) \exp(i(kx-\omega t))\rbrace$ (for example).  The
first order solutions can be found by introducing stream and potential
functions, $\mathbf{v}^{(1)} = \mathrm{curl}  (\Psi \hat{\mathbf{z}})$ and
$\delta \mathbf{u}^{(1)} = \mathrm{curl} (\Phi \hat{\mathbf{z}}) + \nabla
\chi$, and then following the methods detailed in ~\cite{FuShenoyPowers2008}.
Once the first order solutions are obtained, the swimming velocity is
found by solving the second order equations.  The swimming
velocity is determined by $x$-component of the time- and
$x$-averaged velocity and displacement fields.  Note that any overall
time- or $x$-derivative vanishes upon averaging, so
the time-averaged $x$-components of the 2nd order pieces of
Eq.~\ref{gelequations1} and \ref{gelequations2} are 
\begin{eqnarray}
-\eta \partial_y^2 \langle v_x^{(2)} \rangle &=& \mu \partial_y^2 \langle \delta u_x^{(2)} \rangle + \partial_y \langle
  \bar{\sigma}^{{\mathrm n}(2)}_{yx} \rangle\nonumber\\
&&\!\!\!\!\!\!\!\!\!\!\!\!\!\!\!\!\!\!\!\!\!\!\!\! = \Gamma [ V_s - \langle v_x^{(2)}
  \rangle + \langle ( \partial_t \delta \mathbf{u}^{(1)} \cdot
  \nabla)  \delta u_x^{(1)} \rangle ].  
\end{eqnarray}
Here the brackets denote time-and x-averaging and
$\bar{\sigma}^{{\mathrm n}(2)}_{yx}$ includes the 2nd order
contributions to the network stress involving products of two first
order fields.  The solutions to these second order equations are
$\langle v_x^{(2)} \rangle = V_s + c_1 \exp(- \sqrt{\Gamma/\eta} y) +
f(y)$ and $\langle \delta u_x^{(2)} \rangle = -(\eta/\mu) \langle
v_x^{(2)} \rangle + g(y)$, where $f(y)$ is the inhomogeneous solution
satisfying $(\eta \partial_y^2 - \Gamma) f = -\Gamma\langle \partial_t
\delta \mathbf{u}^{(1)} \cdot \nabla \delta u_x^{(1)} \rangle$ and
$g(y)$ is the inhomogeneous solution satisfying $\mu \partial_y g = -
\langle \bar{ \sigma}_{xy}^{{\mathrm n}(2)} \rangle$.  Boundary
conditions at infinity rule out other possible solutions.

These solutions must satisfy the second order boundary conditions,
obtained by expanding Eqs.~\ref{direct} and \ref{solvent} using $\hat {\mathbf{n}} \approx \hat \mathbf{y} -
\partial_x h \hat \mathbf{x}$ and $\hat {\mathbf{t}} \approx \hat \mathbf{x}
+ \partial_x h \hat \mathbf{y}$.  For
direct sheet-network interactions, the relevant BC
are
\begin{eqnarray*}
\mathrm{see~Eqs.~\ref{V2BC}~and~\ref{U2BC}} 
\end{eqnarray*}
\begin{widetext}
\begin{eqnarray}
\langle v_x^{(2)} \rangle &=& - \langle h \partial_y v_x^{(1)} \rangle
\label{V2BC} \\
\left( \begin{array}{c}  \mu \partial_y \langle \delta u_x^{(2)}
  \rangle + \langle \bar{\sigma}_{xy}^{{\mathrm
    n}(2)} \rangle+
  \langle h \partial_y \sigma_{xy}^{{\mathrm n}(1)} \rangle \\
+ \langle \partial_x h \sigma_{yy}^{{\mathrm n}(1)} \rangle  -
\langle \partial_x h \sigma_{xx}^{{\mathrm n}(1)} \rangle \end{array} \right) &=& \Xi \left( \begin{array}{c} V_s +
  \langle \partial_t {\delta \mathbf{u}}^{(1)} \cdot
  \nabla \delta u^{(1)}_x \rangle \\ + \langle h \partial_y
  \partial_t {\delta
  u_x}^{(1)} \rangle  + \langle \partial_t {\delta u_y}^{(1)} \partial_x
  h \rangle
   - \langle \partial_t h \partial_x h \rangle  \end{array}\right) \label{U2BC},
\end{eqnarray}
\end{widetext}
with all quantities evaluated at $y=0$.  Eqs.~\ref{V2BC} and
\ref{U2BC} also apply for solvent-mediated BC,
provided that we set $\Xi=0$ in Eq.~\ref{U2BC}.  In Eq.~\ref{U2BC}, note that all
pieces involving $\langle \bar{ \sigma}_{xy}^{{\mathrm n}(2)} \rangle$
cancel due to the form of $\langle \delta u_x^{(2)} \rangle$.
Therefore nonlinearities in the constitutive relation for the polymer
stress (i.e., $\bar{\bm{\sigma}}^{\mathrm n}$) do not affect the
swimming velocity at second order.  Similar cancellations help to show
that these solutions satisfy overall force balance on the swimmer.

Eliminating $c_1$ from
Eqs.~\ref{V2BC} and \ref{U2BC} yields $V_s$ in terms of
first order quantities.  For direct
network-sheet BC,
\begin{eqnarray*}
\mathrm{see~Eq.~\ref{swimspeed}}
\end{eqnarray*}
\begin{widetext}
\begin{equation}
V_s = - {\left( 1+ \frac{\sqrt{\Gamma \eta}}{\Xi} \right)^{-1}}{\left( \begin{array}{c} \left( -\langle \partial_t h \partial_x h \rangle +  \langle h \partial_y
  \partial_t {\delta
  u_x}^{(1)} \rangle  + \langle \partial_t {\delta u_y}^{(1)} \partial_x
  h \rangle +  \langle \partial_t {\delta \mathbf{u}}^{(1)} \cdot
  \nabla \delta u^{(1)}_x \rangle \right)\\ + \frac{\sqrt{\Gamma \eta}}{\Xi} \left( \langle h
  \partial_y v_x^{(1)} \rangle + f(0) \right) + \frac{\eta}{\Xi}
  \partial_y f(0) - \frac{1}{\Xi}\left( \langle h \partial_y
  \sigma_{xy}^{{\mathrm n}(1)} \rangle  +  \langle \partial_x h (\sigma_{yy}^{{\mathrm n}(1)}
  - \sigma_{xx}^{{\mathrm n}(1)}) \rangle \right)  \end{array} \right)
  }. \label{swimspeed}
\end{equation}  
\end{widetext}
The fourth term on the right hand side, $\langle \partial_t {\delta
  \mathbf{u}}^{(1)} \cdot \nabla \delta u^{(1)}_x \rangle $, arises from the convective nonlinearity.  
For the solvent-mediated BC, the swimming velocity is
given by Eq.~\ref{swimspeed} with $\Xi=0$.  However, the
swimming velocity with solvent-mediated BC is not
just the swimming velocity with direct BC for
$\Xi=0$, since the first-order solutions differ.

For an incompressible network ($\lambda \rightarrow \infty$) the swimming velocity has a simple form
when the convective nonlinearity is included:   
\begin{equation}
V_s = b^2 \omega k \frac{\sqrt{\Gamma \eta} [4 +
6 \sqrt{\Gamma/(\eta k^2)} +\Gamma/(\eta k^2)]}{(\Xi + \sqrt{\Gamma
    \eta}) [2+ \sqrt{\Gamma/(\eta k^2)}]^3}.
\end{equation}
For an incompressible network the swimming
velocity is the same as in a Newtonian fluid, $V_N = b^2
\omega k/2$, when the convective nonlinearity is not included.

\textbf{Results and discussion}.  Our solutions reveal that the
heterogeneous nature of gels leads to behavior not captured
by modeling viscoelastic media as single-phase fluids.  This is
apparent even in the first order solutions.  In a viscoelastic fluid,
the first order solutions are the same as in a Newtonian fluid,
while in our solutions, once the network becomes compressible,
or once one uses solvent-mediated sheet-network BC, there is relative
motion between the solvent and network and both phases move
differently from a Newtonian fluid.

\begin{figure}
\includegraphics[width=8.5cm]{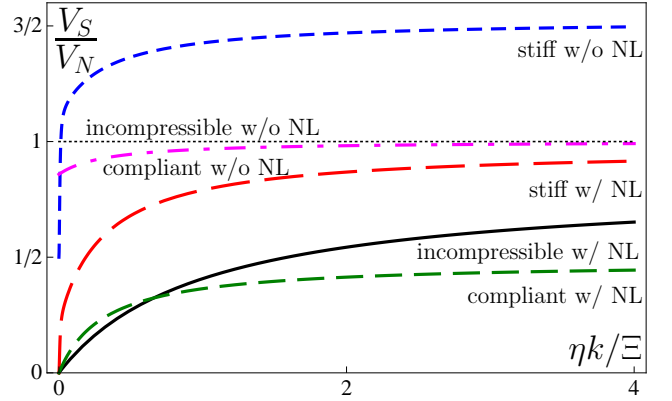}
\caption{
  Swimming speed normalized by Newtonian swimming speed, $V_s/V_N$, as a
  function of $\eta k/\Xi$.   Only when convective 
  nonlinearities are included does the
  swimming velocity vanish for no-slip BC $(\eta k/\Xi \rightarrow 0)$.  
  All curves are for direct network-sheet BC.  
  Thin-black (dots) and thick-black (solid): incompressible network with $\Gamma/(\eta k^2)=1$.  
  Red (long dashes): $\eta \omega/\mu = 0.01$, $\lambda=\mu$,
  $\Gamma/(\eta k^2)=10$.   
  Magenta (dash-dots): $\eta \omega /\mu = 10$, $\lambda=\mu$,
  $\Gamma/(\eta k^2)=10$.   
  Blue (short dashes): $\eta \omega /\mu = 0.01$, $\lambda =\mu$,
  $\Gamma/(\eta k^2)=10$.  
  Green (medium dashes): $\eta \omega /\mu = 10$, $\lambda =\mu$,
  $\Gamma/(\eta k^2)=10$.  
  Thick-black, Red, and Green curves include the convective
  nonlinearity; thin-black, magenta, and blue curves do not.
}
\label{NL}
\end{figure}

Turning to the swimming velocities, we see the importance of correctly
accounting for nonlinearities.  As in previous studies of the role of
nonlinearities in single-phase viscoelastic fluids~\cite{Lauga2007,
  FuPowersWolgemuth2007, FuWolgemuthPowers2008Scallop}, for small
amplitude swimming strokes the swimming speed is a second order
effect, and so a consistent description must take into account any
nonlinearities in constitutive laws.  In the case of a gel described
using a two-fluid model, we have shown that for both direct and
solvent-mediated sheet-network BC, nonlinearities in the constitutive
law for the elastic phase do not affect the swimming velocity.
However, the {\em geometric} convective nonlinearity must be taken
into account to obtain physically meaningful swimming velocities.
Fig.~\ref{NL} shows the swimming velocity as a function of inverse
friction $1/\Xi$.  Previously we argued that no-slip sheet-network BC
prevent a sheet from swimming, which implies that the swimming
velocity should vanish as $1/\Xi \rightarrow 0$, (the no-slip limit).
The figure shows that this expected behavior is only obtained when the
convective nonlinearities are included.  This holds for incompressible
networks ($\lambda \rightarrow \infty$) as well as stiff (small
$\eta\omega/\mu$) and compliant (large $\eta\omega/\mu$) compressible
networks.  In the rest of this paper we always include the convective
nonlinearity.

\begin{figure}
\includegraphics[width=8.5cm]{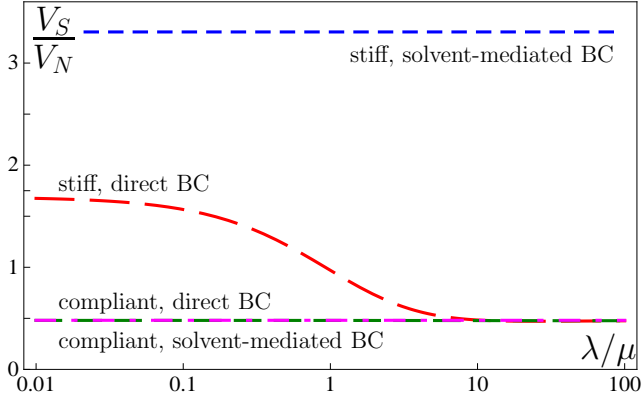}
\caption{Swimming speed normalized by Newtonian swimming speed, $V_s/V_N$, as a
  function of incompressibility, $\lambda/\mu$.  
  Compressibility effects can enhance swimming speeds for direct
  network-sheet BC.  Swimming speed is enhanced
  relative to the Newtonian case for stiff
  networks or low frequencies (small $\eta\omega/\mu$).  
  Red and green curves
  are for direct network-sheet BC and $\Xi=0$; blue and magenta curves are
  for solvent-mediated network-sheet BC. 
  Red (long dashes): $\eta \omega/\mu =0.01$, $\Gamma/(\eta k^2)=10$.  
  Green (medium dashes): $\eta \omega/\mu = 10$, $\Gamma/(\eta k^2)=10$.
  Blue (short dashes): $\eta \omega/\mu = 0.01$, $\Gamma/(\eta k^2)=10$.  
  Magenta (dash-dots): $\eta \omega/\mu = 10$, $\Gamma/(\eta k^2)=10$.  
}\label{compressible}
\end{figure}

Compressibility effects are a possible mechanism for swimming speed
enhancement.  Fig.~\ref{compressible} show the swimming speed as a
function of Lam\'e coefficient $\lambda$, a measure of
incompressibility (the bulk modulus of compression is $2 \mu +
\lambda$).  For direct sheet-network BC, the
strongest enhancement of swimming speed occurs for a material with
large shear modulus ($\mu \rightarrow \infty$) but which is maximally
compressible ($\lambda = 0$), and for a network which slips freely
past the swimming sheet ($\Xi=0$).  Compressibility does not strongly
affect the swimming speed for solvent-mediated sheet-network BC. Swimming speeds can be faster than in Newtonian fluids for
stiff networks/low frequencies (small $\eta\omega/\mu$), but not for
compliant networks/high frequencies (large $\eta\omega/\mu$),
reflecting a broader distinction which we address next.
  
\begin{figure}
\includegraphics[width=8.5cm]{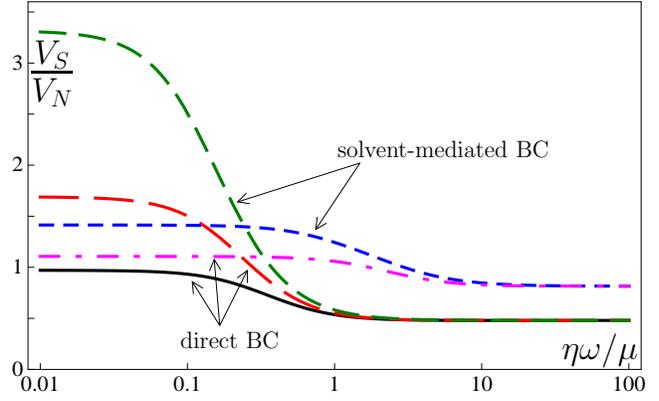}
\caption{ 
  Swimming speed normalized by Newtonian swimming speed, $V_s/V_N$, as a
  function of $\eta \omega/\mu$.  
  There is a transition at $\eta \omega /\mu \sim \eta k^2/\Gamma$
  between fast swimming (stiff network/low frequencies, small $\eta
  \omega/\mu$) and slow swimming (compliant network/high frequences,
  large $\eta\omega/\mu$).    
  Black, red and magenta curves are for
  direct network-sheet BC and $\Xi=0$; blue and green curves are
  for solvent-mediated network-sheet BC. 
  Black (solid): $\lambda=\mu$, $\Gamma/(\eta k^2)=10$.
  Red (long dashes): $\lambda=0$, $\Gamma/(\eta k^2)=10$.  
  Magenta (dash-dots): $\lambda = 0$, $\Gamma/(\eta k^2)=1$.  
  Green (medium dashes): $\lambda=0$, $\Gamma/(\eta k^2)=10$.
  Blue (short dashes): $\lambda = 0$, $\Gamma/(\eta k^2)=1$.  
}\label{stiff}
\end{figure}

Fig.~\ref{stiff} plots the swimming velocity as a function of $\eta
\omega/\mu$.  There is a transition between two regimes, one for
flexible networks or high frequency beating (small $\eta\omega/\mu$),
leading to a smaller swimming velocity; and one for rigid networks or
low frequency beating (large $\eta\omega/\mu$), leading to larger
swimming velocities.  The swimming speed enhancement is larger for
larger values of $\Gamma$, and is larger for solvent-mediated
network-sheet BC than for direct network-sheet
BC.  Physically, the transition between the two
regimes corresponds to whether the network does not deform (stiff
network/low frequencies) or whether the network deforms to move
together with the solvent (compliant network/high frequencies).  Since
the network deforms due to drag forces from the solvent, one can
predict that the transition between the two regimes occurs when the
force/volume due to drag forces, $\sim \omega u \Gamma$, becomes
comparable to the force/volume due to elasticity, $\sim \mu u k^2$,
i.e. when $\eta \omega/\mu \sim \eta k^2/\Gamma$, which can be
confirmed by comparing the curves in Fig.~\ref{stiff}.

\begin{figure}
\includegraphics[width=8.5cm]{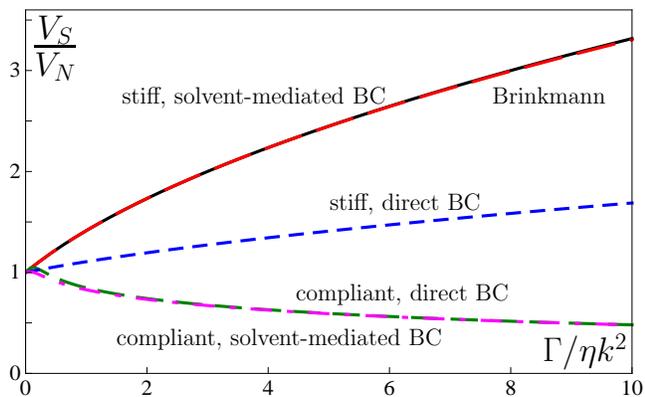}
\caption{ 
Swimming speed normalized by Newtonian swimming speed, $V_s/V_N$, as a
  function of drag coupling $\Gamma/\eta k$.  The swimming speed
  is the same as in a Brinkmann fluid (black solid curve) for solvent-mediated
  network-sheet BC when the network does
  not deform (small $\eta\omega/\mu$).   
  Blue and magenta curves are for
  direct network-sheet BC, $\Xi=0$, and $\lambda=0$;
  red and green curves are for solvent-mediated network-sheet BC and
  $\lambda=0$.  
  Red (long dashes): $\eta\omega/\mu=0.01$.  
  Green (medium dashes): $\eta\omega/\mu=10$.
  Blue (short dashes): $\eta\omega/\mu=0.01$.  
  Magenta (dash-dots): $\eta\omega/\mu=10$.  
}\label{gamma}
\end{figure}

We can compare our results to those for the single-phase Brinkmann
fluid~\cite{SiddiquiAnsari2003, Leshansky2009}, which obeys $\eta \nabla^2
\mathbf{v} - \nabla p = \Gamma^B \mathbf{v}$ and has swimming speed
$V_B = V_N \sqrt{1 + \Gamma^B/(\eta k^2)}$.  In the Brinkmann fluid the
elastic phase is not a dynamic degree of freedom, but instead enters
as an inert background retarding solvent motion.  One expects that the
two-fluid model behaves like a Brinkmann fluid when the network does
not deform (i.e., $\eta \omega/\mu \ll \eta k^2/\Gamma$) and if
$\Gamma^B = \Gamma$.  This expectation is borne out in
Fig.~\ref{gamma}, but only for solvent-mediated BC.  
When the network deforms (compliant networks/high
frequencies) the network-sheet BC do not
greatly affect the swimming speed (magenta and green curves).

Our results highlight the importance of understanding the proper
boundary conditions for swimmers in gels.  The boundary conditions
need not be the same for all swimmers; they may depend on factors
including swimmer morphology, surface biochemistry, or gel structure.
However, one way to estimate whether a swimmer has direct interaction
with the polymer network in a gel is to compare the swimmer size with
the mesh size of the network.  For example, sperm swimming in cervical
mucus have head sizes of 3-5 $\mu$m.  The mesh size of the network of
mucin fibers in cervical mucus varies strongly with time during the
menstrual cycle, with mesh sizes of up to 25 $\mu$m around ovulation,
and as small as 200 nm at other times~\cite{Ruttlant_etal2005,
  Ruttlant_etal2001}.  Thus, during ovulation, when the sperm may be
able to swim through the mucus while rarely encountering mucin fibers,
solvent-mediated network-sheet BC likely apply.  When the mesh sizes
are small, direct network-sheet BC are more appropriate.  Therefore,
both types of boundary conditions described in this work may be
relevant in biological situations.  One important lesson is that even
for an effective medium model of a heterogeneous material (such as our
two-fluid model), microscopic heterogeneity will have strong consequences for
the boundary conditions and corresponding swimming properties, since
different boundary conditions lead to nearly 100\% changes in swimming
speeds in the stiff network limit.  

Note that for the particular case of sperm in cervical mucus, the
sperm are likely to be in the compliant network/high frequency phase:
for well-hydrated mucus (near ovulation) when the mesh size is
$\approx 10 \; \mu$m, the shear modulus is $\approx 1$ Pa, and
viscosity is $\approx 0.4$ Pa s; while farther from ovulation with
mesh size $\approx 0.5 \; \mu$m, the shear modulus is $\approx 4$ Pa
and viscosity is $\approx 20$ Pa s~\cite{TamKatzBerger1980}.
Estimating $\Gamma \approx \eta/\xi^2$, where $\xi$ is the mesh size,
and using a wavelength of sperm flagellar beating ($\approx 10 \;
\mu$m) to determine $k$, we estimate that near ovulation
$\eta\omega/\mu \approx 10$ and $\Gamma/(\eta k^2) \approx 0.03$,
while away from ovulation $\eta\omega/\mu \approx 100$ and
$\Gamma/(\eta k^2) \approx 10$.  Although for sperm the swimming speed
enhancement seen in the stiff network/low frequency regime is unlikely
to be realized in cervical mucus, for other swimmers in gels it may
still be an important effect.

As we were completing this letter, we learned of the work of
H. Wada~\cite{Wada2010}, who has independently studied a closely related
problem. 

We are grateful to Charles Wolgemuth for many insightful
conversations. This work was supported in part by National Science
Foundation Grants Nos. CMMI-0825185 (VBS), DMS-0615919 (TRP), CBET-0854108
(TRP). HF and TRP thank the Aspen Center for Physics, where some of
this work was done.  We are grateful to H. Wada for sharing his work
with us before it was published.


\end{document}